\documentclass[conference]{IEEEtran}

\usepackage[tight,footnotesize]{subfigure}
\usepackage{url}
\usepackage{listings}
\usepackage{xcolor}

\colorlet{punct}{red!60!black}
\definecolor{background}{HTML}{EEEEEE}
\definecolor{delim}{RGB}{20,105,176}
\colorlet{numb}{magenta!60!black}

\lstdefinelanguage{json}{
    numbers=left,
    stepnumber=1,
    numbersep=8pt,
    showstringspaces=false,
    breaklines=true,
    frame=lines,
    literate=
     *{0}{{{\color{numb}0}}}{1}
      {1}{{{\color{numb}1}}}{1}
      {2}{{{\color{numb}2}}}{1}
      {3}{{{\color{numb}3}}}{1}
      {4}{{{\color{numb}4}}}{1}
      {5}{{{\color{numb}5}}}{1}
      {6}{{{\color{numb}6}}}{1}
      {7}{{{\color{numb}7}}}{1}
      {8}{{{\color{numb}8}}}{1}
      {9}{{{\color{numb}9}}}{1}
      {:}{{{\color{punct}{:}}}}{1}
      {,}{{{\color{punct}{,}}}}{1}
      {\{}{{{\color{delim}{\{}}}}{1}
      {\}}{{{\color{delim}{\}}}}}{1}
      {[}{{{\color{delim}{[}}}}{1}
      {]}{{{\color{delim}{]}}}}{1},
}
\makeatletter
\def\lst@makecaption{%
  \def\@captype{table}%
  \@makecaption
}
\makeatother

\usepackage{tikz}
\usepackage{xcolor}
\usepackage{color}
\usepackage{pgfplots}

\usetikzlibrary{
	arrows.meta,
	calc,
	colorbrewer,
	external,
	fit,
	matrix,
	positioning,
	shapes,
	shapes.misc,
	backgrounds
}

\begin{document}

\title{MONICA in Hamburg: Towards \\ Large-Scale IoT Deployments in a Smart City}

\author{
	\IEEEauthorblockN{
		Sebastian Meiling\IEEEauthorrefmark{1},
		Dorothea Purnomo\IEEEauthorrefmark{3},
		Julia-Ann Shiraishi\IEEEauthorrefmark{2},
		Michael Fischer\IEEEauthorrefmark{3}, and
		Thomas C. Schmidt\IEEEauthorrefmark{1}
	}
	\vspace{1ex}
	\IEEEauthorblockA{
		\IEEEauthorrefmark{1}\{sebastian.meiling, t.schmidt\}@haw-hamburg.de, 
		\IEEEauthorrefmark{2}julie-ann.shiraishi@sk.hamburg.de,\\
		\IEEEauthorrefmark{3}\{dorothea.purnomo, michael.fischer1\}@gv.hamburg.de}
	\vspace{1ex}
	\IEEEauthorblockA{
		\IEEEauthorrefmark{1}iNET RG, Hamburg University of Applied Sciences, Germany}
	\IEEEauthorblockA{
		\IEEEauthorrefmark{2}Senate Chancellery, Free and Hanseatic City Hamburg, Germany}
	\IEEEauthorblockA{
		\IEEEauthorrefmark{3}Agency for Geoinformation and Surveying, Free and Hanseatic City Hamburg, Germany}
}
\maketitle

\begin{abstract}
Modern cities and metropolitan areas all over the world face new management challenges in the 21st century primarily due to increasing demands on living standards by the urban population.
These challenges range from climate change, pollution, transportation, and citizen engagement, to urban planning, and security threats.
The primary goal of a Smart City is to counteract these problems and mitigate their effects by means of modern ICT to improve urban administration and infrastructure.
Key ideas are to utilise network communication to inter-connect public authorities; but also to deploy and integrate numerous sensors and actuators throughout the city infrastructure -- which is also widely known as the Internet of Things (IoT).
Thus, IoT technologies will be an integral part and key enabler to achieve many objectives of the Smart City vision.

The contributions of this paper are as follows.
We first examine a number of IoT platforms, technologies and network standards that can help to foster a Smart City environment.
Second, we introduce the EU project MONICA which aims for demonstration of large-scale IoT deployments at public, inner-city events and give an overview on its IoT platform architecture. 
And third, we provide a case-study report on SmartCity activities by the City of Hamburg and provide insights on recent (on-going) field tests of a vertically integrated, end-to-end IoT sensor application.
\end{abstract}

\section{Introduction}
\label{sec:intro}

A general problem of modern cities in the twenty-first century is the rapid increase of the urban populations~\cite{np-cscdt-11}, this urbanisation is a world wide phenomenon. 
The specific challenges range from physical: air pollution and waste, health issues, lack of transport and living space; to economic and societal: unemployment, price instability, resource scarcity and integration.
To that end, Smart City projects need to find advanced and innovative solutions, e.g., for urban planning, that make future cities sustainably liveable.

One of the key success factors of Smart City missions is Information and Communication Technology (ICT): smart computing refers to a new generation of integrated hardware, software and network technologies that provide IT-systems with real-time awareness of the physical world and advanced analytics to help people make more intelligent decisions about alternatives and actions that will optimise processes and results~\cite{cnwgm-uscif-12}.
Smart Governance, for instance, represents a collection of technologies, people, policies, practices, resources, social norms and information that interact to support city governing activities.
In a smart city, the integration of information and communication technologies into a city's different technical systems and infrastructures is the basis for innovative solutions in the fields of energy, administration, health, mobility and security.

A key idea of the Internet of Things (IoT)~\cite{aim-its-10} is to enhance the real-world with inter-connected devices, i.e., sensors and actuators, to enable new services and applications.
As such IoT technologies generally fit into the Smart City vision and will complement existing urban infrastructures to enable innovative solutions to address the aforementioned challenges.
 
The paper is structured as follows: in Section~\ref{sec:background} we review a variety of Smart City  enabling IoT platforms and technologies. 
Section~\ref{sec:monica} introduces the EU project MONICA, its objectives and the architecture of an open IoT platform for large-scale deployments. 
Section \ref{sec:hamburg} gives an overview on Smart City efforts and the Urban Platform ---the central open data hub--- of the City of Hamburg.
In Section \ref{sec:trials} we report on IoT deployment and integration tests performed in Hamburg, which also is ongoing work.

\section{Background and Related Work}
\label{sec:background}

The Internet of Things (IoT) extends the traditional Internet into the world of embedded, resource constraint devices utilising machine-to-machine communication based on wireless, low-power radio standards.
Hence, the IoT requires dedicated architectures and technologies specifically designed to be compliant with  such constraint environments, but that also allow for scalability and robustness.

\subsection{IoT Architectures and Platforms}
The goal of the \emph{IoT-European Platforms Initiative} (IoT-EPI) is to build a sustainable IoT-ecosystem in Europe and aims for interoperability of platforms.
For instance, \emph{Open-IoT}\footnote{\url{http://www.openiot.eu}} which provides an open source middleware to enable cloud-based ``Sensing-as-a-Service''.
FIWARE is an EU originated open source platform based on standardised APIs and software components to implement vertically integrated, smart IoT services.
The \emph{FIESTA-IoT}~\cite{gs-ffisi-16} EU project provides middleware infrastructure for IoT experiments and  federates existing IoT platforms and testbeds under a common ontology.

The \emph{AIOTI High-Level Architecture} (HLA)~\cite{aioti-hla-17} is a function model for an IoT architecture composed of three main layers: application, IoT, and network layer.
The HLA describes functions and interfaces for entities within a layer, but also the interaction between layers.
However, the AIOTI HLA does not stipulate details on implementation or deployment.
The Open Geospatial Consortium (OGC) defined the \emph{SensorThings API}~\cite{lhk-ostap-16} which is an open standard and geospatial-enabled framework to interconnect IoT devices, applications and data; addressing syntactic and semantic interoperability.
Besides a RESTful request/response architecture based on HTTP, the API also supports a publish/subscribe architecture using MQTT.
That may yield lower latency and allows for many-many distribution of sensor data similar to IP multicast.
The \emph{OneM2M Functional Architecture} (FA)~\cite{onem2m-fa-16} describes end-to-end services based on functional entities and reference points.
It is primarily service oriented, but remains independent of the underlying network layers.
While it shares the overall layered view of the AIOTI HLA, the oneM2M FA also specifies distinct service functions for device management and dedicated interfaces.
The \emph{Web of Things Architecture} (WoT)~\cite{wot-architecture} is an upcoming standard by the World Wide Web Consortium (W3C) that aims for interoperability across IoT platforms.
The W3C WoT specifies (RESTful) interfaces to enable communication between devices and services in the IoT, independent of implementation and communication technologies.

To implement innovative IoT applications on a wide range of embedded devices and platforms it requires IoT operating systems~\cite{hbpt-osldi-16}.
The iNET research group co-founded and core-develops RIOT~\cite{bhgws-rotoi-13}, a widely popular free open source operating system for the IoT.
\emph{RIOT-OS}\footnote{\url{https://riot-os.org}} is an open platform and ecosystem for tiny devices \cite{bghkl-rosos-18}, very similar to what Linux is for standard computers.
This ultra-lean OS supports most embedded, low-power devices, platforms, and various micro-controller architectures (from 8 to 32-bit).
Its key features are: (a) low memory footprint (few kBs), (b) tick-less scheduling for near realtime and energy efficiency; (c) reduced hardware dependent code combined with high level APIs; complemented by (d) multithreading and (e) a flexible network stack.
RIOT aims to implement all relevant open standards ---with focus on network protocols--- supporting an Internet of Things that is connected, secure, durable, and privacy-friendly~\cite{lkhpg-cwemr-18}.
The development is driven by a growing international community gathering companies, academia, and inventors.

\subsection{IoT Technologies, Standards, and Protocols}
Besides architectures and platforms, there is also a need for network stacks that vertically combine protocols suitable for the IoT.
Over the last years many protocols were developed and standardised that are specifically designed to cope with constraint devices and low-power networks.
On the physical and link layer there are several low-power radio standards such as IEEE 802.15.4, LoRa,  SigFox, and 5G Narrowband-IoT (NB-IoT).
These standards typically provide only limited bandwidth compared for instance to WiFi, i.e., an IEEE 802.15.4 frame has a size of up to 127 bytes.
To transparently inter-connect the IoT with the common Internet, it needs to utilise the Internet Protocol on the network layer.
With hundreds of thousands (even millions) devices in the IoT, IPv6 is the only option, but is also challenging considering 128 bit addresses and a minimal MTU of 1280 bytes.
To that end 6LoWPAN combines several techniques, such as generic header compression and fragmentation to transparently cope with the IPv6 MTU, into an adaption layer to support IPv6 on constraint networks and embedded IoT devices.
To distribute data in large wireless IoT networks it further requires a routing protocol like RPL~\cite{RFC-6550} to allow for multi-hop communication.
On the application layer \emph{CoAP}~\cite{RFC-7252} allows to implement RESTful data transfer and services using a requests-response scheme similar to HTTP.
CoAP supports a variety of data formats (MIME) to encode message payloads, possible formats are plain text, CBOR~\cite{RFC-7049}, JSON, and SenML~\cite{draft-ietf-core-senml}, to name only a few.
Unlike HTTP in the traditional Internet, CoAP uses UDP on the transport layer instead of TCP, which is not suited for constraint IoT networks due to its overhead.

\section{The MONICA Project}
\label{sec:monica}

The MONICA project\footnote{\url{https://www.monica-project.eu}} is part of the IoT European Large-Scale Pilots Programme (LSP)\footnote{\url{https://european-iot-pilots.eu}} and as such aims for large scale demonstration of an IoT ecosystem based on multiple existing and new IoT technologies for Smarter Living.

\subsection{Objectives and Scenarios}
\label{ssec:monica:objectives}

The focus of MONICA is on a key aspect of the European society: the cultural performances in open-air settings which create challenges in terms of crowd safety, security, and noise pollution.
To that end MONICA utilises innovative wearables, sensors and actuators with closed-loop back-end services integrated into an interoperable, cloud-based IoT platform capable of offering a variety of simultaneous, targeted applications to thousands of users.
The MONICA IoT platform will be demonstrated by IoT applications addressing environmental and safety issues associated with large open-air events held in inner cities (e.g., concerts, funfairs or sports matches).
Two main ecosystems will be implemented at large-scale: (a) a Security Ecosystem and (b) an Acoustic Ecosystem.

The first will show how a multitude of innovative applications for managing public security and safety can be seamlessly integrated with numerous IoT sensors and actuators.
Major security and safety challenges at large inner-city events are the handling and mitigation of unforeseen incidents.
This includes personal violence, threats of terrorist attacks, panic scenes, and severe illness of individuals in the middle of a crowd.
The MONICA IoT platform will collect and analyse sensor data to enable decision support systems (DSS) and a common operational picture (COP).
The Acoustic Ecosystem will demonstrate smart applications for managing open-air music performances in urban spaces by seamlessly integrating IoT devices using the MONICA platform.
The system will consist of a series of large and small applications that, in combination, can be used to monitor and manage the sound before, during, and after a performance.
The main objectives are to demonstrate how the IoT platform supports closed loop solutions that address real-life environmental challenges such as the reduction of noise in public spaces close to open-air events.

The two ecosystems are also connected; for instance an acoustic-based module may detect security related events such as gunshot, scream, breaking glasses and approaching vehicles. 
Moreover, the MONICA platform can be integrated with IoT applications and resources from other Smart City platforms (e.g., the Hamburg Urban Platform) and external open data portals.
To than end, MONICA aims for interoperability with other EU research projects and initiatives, such as AIOTI, IoT-EPI, FIWARE, Fiesta-IoT, Open-IoT, CRYSTAL and SOFIA.

\subsection{IoT Platform Architecture}
\label{ssec:monica:arch}

The MONICA architecture comprises of the following subsystems and layers, starting from bottom:
\begin{itemize}
	\item \emph{Device Layer:} includes all fixed and mobile IoT appliances, such as wearables, sensors and actuators.
	\item \emph{Network Layer:} support heterogeneous IoT devices with different access networks standards, e.g., Wi-Fi, cellular (3/4/5G), and low-power WAN (IEEE 802.15.4).
	\item \emph{Edge Layer:} consists of gateways to connect devices with the MONICA backend and processing nodes for real-time data.
	\item \emph{Adaption Layer:} provides technology independent management of resources and uniform mapping of data into standard representations, implemented by SCRAL.
	\item \emph{IoT Platform Connectors:} handles communication and data integration with external IoT platforms, e.g., exposes IoT data according to the OneM2M standard.
	\item \emph{Middleware:} provides storage and directory service for registered resources; is implemented by LinkSmart.
	\item \emph{Services Layer:} provides service-specific data processing and sensor fusion, a common knowledge base, and decision support tools.
	\item \emph{MONICA APIs Layer:} comprises of distinct APIs for public and professional (security) applications.
	\item \emph{Cyber Security and Privacy Framework:} enables and ensures trust-based communication, secure data flow and storage across all components of the MONICA platform.
	\item \emph{Deployment and Monitoring Tools:} check and verify deployment of MONICA platform components periodically; also enable for performance measurements. 
\end{itemize}

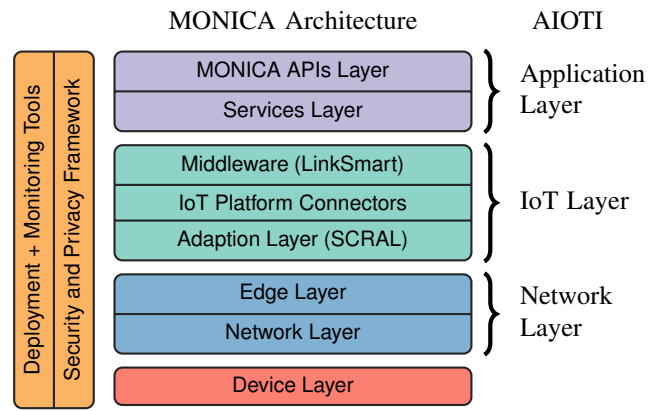
\begin{figure}
	\centering
\definecolor{DarkGrey}{RGB}{37,37,37}		
\definecolor{LiteGrey}{RGB}{217,217,217} 	
\definecolor{IoT}{RGB}{141,211,199} 			
\definecolor{Net}{RGB}{128,177,211} 			
\definecolor{Else}{RGB}{253,180,98}			
\definecolor{App}{RGB}{190,186,218}			
\definecolor{Dev}{RGB}{251,128,114}			

\tikzset{
	layer/.style = {
		draw=DarkGrey,
		rounded corners=3pt, 
		rectangle, 
		thick, 
		minimum width=30ex,
		minimum height=3ex,
		font=\footnotesize\sffamily,
	},
	descr/.style = {
		text width = 10ex,
		xshift = 8ex,
	},
	app/.style	= {fill=App},
	iot/.style	= {fill=IoT},
	net/.style	= {fill=Net},
	dev/.style	= {fill=Dev},
	tool/.style	= {fill=Else},
	every node/.style = {
		node distance=1ex,
	},
}
\tikzsetnextfilename{monica}
\begin{tikzpicture}
	\node[minimum width=32ex] (monica) {MONICA Architecture};
	\node[minimum width=12ex, right=of monica] (aioti) {AIOTI};
	\node[layer, app, rectangle split, rectangle split parts=2, below=of monica] (app) 
		{\nodepart[text centered]{one}MONICA APIs Layer
		 \nodepart[text centered]{two}Services Layer};
	\node[layer, iot, rectangle split, rectangle split parts=3, below=of app] (iot) 
		{\nodepart[text centered]{one}Middleware (LinkSmart)
		 \nodepart[text centered]{two}IoT Platform Connectors
		 \nodepart[text centered]{three}Adaption Layer (SCRAL)};
	\node[layer, net, rectangle split, rectangle split parts=2,below=of iot] (net) 
		{\nodepart[text centered]{one}Edge Layer
		 \nodepart[text centered]{two}Network Layer};
	\node[layer, dev, below=of net] (devices) {Device Layer};
	\node[layer, tool, rectangle split, rectangle split parts=2,left=5ex of app.north west, rotate=90] (tools) 
		{\nodepart[text centered]{two}Security and Privacy Framework
		 \nodepart[text centered]{one}Deployment + Monitoring Tools};
	\draw[ultra thick,decorate,decoration={brace,amplitude=1ex}] ([xshift=1ex]app.north east) -- ([xshift=1ex]app.south east) node [descr,midway] {Application Layer};
	\draw[ultra thick,decorate,decoration={brace,amplitude=1ex}] ([xshift=1ex]iot.north east) -- ([xshift=1ex]iot.south east) node [descr,midway] {IoT Layer};
	\draw[ultra thick,decorate,decoration={brace,amplitude=1ex}] ([xshift=1ex]net.north east) -- ([xshift=1ex]net.south east) node [descr,midway] {Network Layer};
\end{tikzpicture}
	\caption{Overview on layers of the MONICA Architecture and corresponding layers in the AIOTI high-level architecture specification.}
	\label{fig:monica}
\end{figure}

Figure~\ref{sec:intro} shows the mapping of layers and components of the MONICA platform and the corresponding layers in the AIOTI high level architecture.
While the latter is independent of any implementation, the MONICA architecture on the other hand defines distinct software components that implement certain functions of the platform, e.g., the LinkSmart middleware that handles data storage, resource catalogues, and data provision for services above.
Moreover, it goes beyond the AIOTI specification by defining additional vertical layers, to ensure security and privacy but also to allow for monitoring.
And allows for dedicated IoT platform connectors to integrate the MONICA IoT environment with external third parties and services, for instance with existing platforms at a MONICA pilot side (see also~\ref{ssec:hhplatform}).

\section{Hamburg -- Smart City and MONICA Pilot Site}
\label{sec:hamburg}
With more than 1.8 million citizens the Free and Hanseatic City Hamburg is the second largest city in Germany, and the heart of a flourishing metropolitan area in northern Germany.
Within the MONICA project Hamburg and its local partners participate as a pilot site to demonstrate large-scale IoT deployments at popular inner-city events, namely the DOM funfair and the annual Port Anniversary.

\subsection{Overview on Smart City Activities}
\label{ssec:hhoverview}

It has been recognised that the digital transformation cannot be outsourced or delegated. 
Instead, the transformation process has to be exemplified by the city, the administrative body itself.
The main goals of Hamburg's development strategy are to improve the quality of life and to drive economic growth and the city's economic attractiveness.
To achieve these and to address many of the city's challenges, Hamburg pursues a clear strategy that is based on government-driven policies; one of these is the \emph{2015 Digital City Strategy}.
The main objective of this is to support and promote innovation that is based on technology and digitisation.
In the framework of this strategy, the exchange of data between municipal bodies, city systems, and open data users plays a significant role, which is reflected by the Hamburg Urban Platform (cf.~\ref{ssec:hhplatform}).

From a Smart City perspective there are two focus areas in Hamburg, namely:
(a) \emph{Intelligent Transport Systems} (ITS)\footnote{The city of Hamburg will host the ITS World Congress in 2021 \\ \phantom{see} \url{http://www.its2021.hamburg}}, which takes into account that environmental pollution, climate change and sustainability initiatives are inherently linked to traffic.
Hence, green traffic and ITS are among of the main topics of interest around which the City of Hamburg plans its future.
Urban mobility includes concepts such as smart parking, intelligent traffic management, integrated transport.
And (b) \emph{e-governance and citizen services}, which include measures to improve public information, electronic service delivery, citizen engagement, video crime monitoring, and public security.
The goal is to make the city's administration more transparent, cost effective and citizen-friendly.
As of 2018, Hamburg established the Department for IT and Digitisation which is dedicated to these efforts and combines formerly segregated topics in the field of e-governance and Smart City under one authority.

\subsection{The Hamburg Urban Platform}
\label{ssec:hhplatform}

The Urban Platform of the City of Hamburg is a conceptual approach which follows the EIP-SCC initiative of Urban Platforms~\cite{hs-radp-16} and the DIN~SPEC~91357.
It implements a logical architecture that integrates data flows within and across city systems and exploits modern technologies (sensors, cloud services, mobile devices, analytics, etc.).
The Urban Platform enables the City of Hamburg to break up the monolithic data silos while shifting from fragmented operations to inclusive, predictive, and efficient operations and novel ways of engaging and serving city stakeholders.
It contains open and non-open data of different public authorities as well as third parties.
It holds geospatial information on urban data to several categories, e.g., education, culture, urban development and planning, environment, traffic.
These are distributed via standardised web services ---with APIs and data models conforming to OGC specifications, e.g., WFS, WMS and GML--- for viewing, downloading and processing of data.

\subsubsection{Dataset}
\begin{figure}[t]
	\centering
	\input{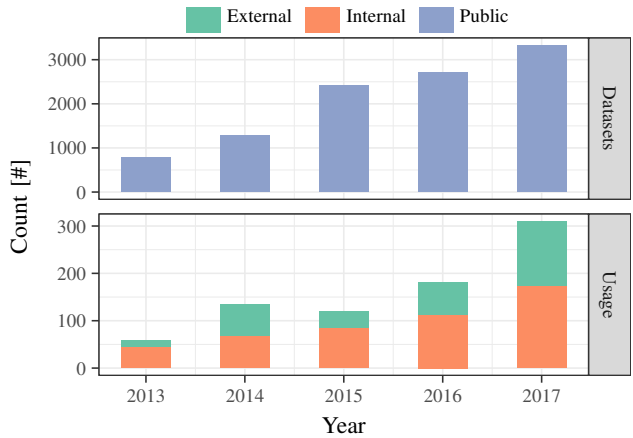}
	\caption{Available public datasets and their usage (in terms of millions web requests per year) over the last 5 years. Usage is broken down into \emph{internal} and \emph{external}, i.e., web requests from city authorities and third parties, respectively.}
	\label{fig:hhup_stats}
\end{figure}
As of 2017 the Urban Platform provides more than 3300 datasets, 93 applications in form of geo portals and more than 400 distinct services which receive more than 310 million requests per year ($\approx$ 850.000 request per day).
Fig.~\ref{fig:hhup_stats} shows the growth in number of datasets and user requests (in millions per year) over the last 5 years.
The urban platform also started to provide near real time information from sensor data e.g. occupation of charging stations for electro mobility, availability of city bikes at the specific bike stations, and availability of parking slots on parking decks.
\subsubsection{Architecture}
\begin{figure}[t]
	\centering
\definecolor{DarkGrey}{RGB}{37,37,37}		
\definecolor{LiteGrey}{RGB}{217,217,217} 	
\definecolor{Auth}{RGB}{141,211,199} 		
\definecolor{Unit}{RGB}{128,177,211} 		
\definecolor{App}{RGB}{253,180,98}			
\definecolor{Other}{RGB}{190,186,218}		
\definecolor{Device}{RGB}{251,128,114}		
\definecolor{Lemon}{RGB}{255,237,111}		

\tikzset{
	layer/.style = {
		draw=DarkGrey, 
		rectangle, 
		thick,
		font=\sffamily\footnotesize\bfseries,
		text=DarkGrey,
		minimum width=.8\columnwidth,
		minimum height=6ex,
		outer sep=0pt,
		inner sep=0pt,
	},
	conn/.style = {
		black,
		rectangle,
		rounded corners=2pt, 
		fill=LiteGrey,
		minimum width=.5\columnwidth,
		minimum height=3ex,
		font=\sffamily\scriptsize,
	},
	app/.style = {
		draw=App,
		ultra thick,
		fill=App!75, 
		minimum height=3ex,
		minimum width=10ex,
		rectangle,
		rounded corners=3pt,
		text width=9ex,
		font=\sffamily\scriptsize
	},
	dev/.style = {
		draw=Device,
		ultra thick,
		fill=Device!75, 
		minimum height=3ex,
		minimum width=8ex,
		rectangle,
		rounded corners=3pt,
		text width=7ex,
		font=\sffamily\scriptsize
	},
	mod/.style = {
		draw=DarkGrey,
		fill=Lemon!75, 
		rectangle,
		rounded corners=1pt,
		font=\sffamily\tiny\bfseries
	},
	mgmt/.style = {
		draw=Auth,
		ultra thick,
		fill=Auth!75, 
		minimum height=3ex,
		minimum width=8ex,
		rectangle,
		rounded corners=3pt,
		text width=10ex,
		font=\sffamily\scriptsize
	},
}
\tikzsetnextfilename{hhup_arch}
\begin{tikzpicture}
	\node [app] (a1) {eGovernance Applications}; 
	\node [app, right=1ex of a1] (a2) {3rd Parties Applications}; 
	\node [app, right=1ex of a2] (a3) {Open Data Portal}; 
	\node [layer, fit={(a1) (a2) (a3)}] (l_app) {};
	\node [layer, fill=Unit, below=3ex of l_app] (l_svc) {Web Services Modules};
	\node [layer, fill=Unit, below=.5ex of l_svc] (l_dat) {Data Warehouse Storage};
	\node [layer, fill=Unit, below=.5ex of l_dat] (l_mod) {Adaptors};
	\node [layer, below=3ex of l_mod, minimum height=5ex,] (l_sys) {};
	\node [conn, anchor=north, below=0ex of l_app.south] (api) {Connectivity -- Standard APIs};
	\node [conn, anchor=south, above=0ex of l_sys.north] (net) {IoT Access Networks};
	\node [mod, above left=2ex and 17ex of net.north] (m1) {JSON};
	\node [mod, above left=1ex and 12ex of net.north] (m2) {XML};
	\node [mod, above left=2.5ex and 7ex of net.north]  (m3) {CSV};
	\node [mod, below left=3.25ex and 18ex of api.south] (s1) {data};
	\node [mod, below left=.4ex and 13ex of api.south] (s2) {processing};
	\node [mod, below left=3.4ex and 11ex of api.south]  (s3) {sensors};
	\node [mgmt, below right=2pt and 10ex of api.south] (am) {Authorization Management};
	\node [mgmt, above right=2pt and 10ex of net.north] (sm) {Sensor Management};
	\node [dev, below left=1ex and 12ex of net.south] (d1) {System A};
	\node [dev, right=1ex of d1] (d2) {System B};
	\node [below=2ex of net.south, font=\bfseries] (dots) {...};
	\node [dev, below right=1ex and 12ex of net.south] (d3) {Sensor B};
	\node [dev, left=1ex of d3] (d4) {Sensor A}; 
	\draw[ultra thick,decorate,decoration={brace,amplitude=1ex}] ([xshift=1ex]l_svc.north east) -- ([xshift=1ex]l_mod.south east) node [xshift=3ex,midway,rotate=90] {Urban Platform Core};
\end{tikzpicture}
	\caption{Overview of the architecture of the Hamburg Urban Platform.}
	\label{fig:hhup_arch}
\end{figure}
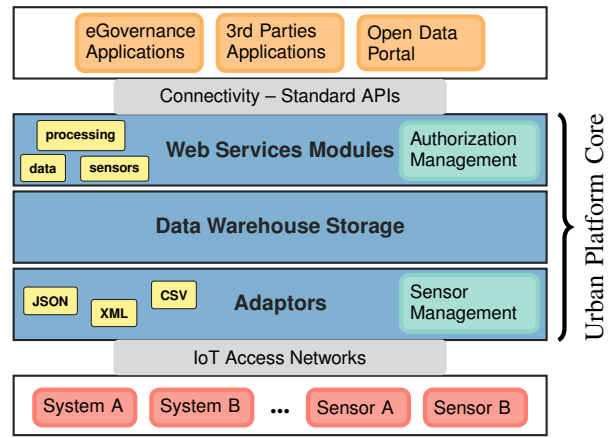
The open Hamburg Urban Platform follows the common architectural framework developed in the EU Project \emph{Espresso}\footnote{\url{http://espresso-project.eu}} and a system-of-systems approach to allow integration with heterogeneous systems or platforms.
The core of the data management layer (see Fig.~\ref{fig:hhup_arch}) is divided in five modules: Data Web Services, Metadata Web Services, Processing Web Services, Data Analytics and Sensor Web Services. 
While the former four are fully deployed (and extended regularly), the latter is under development.
The five modules are substantiated by the Data Warehouse where all data is stored and extracted for the different services. 

\subsubsection{Interoperability and open APIs}
Interoperability is crucial for the success of any Smart City infrastructure in general, and hence for the Urban Platform of the City of Hamburg itself.
The Urban Platform provides easy access to existing data using standardised data formats and APIs with an open design ensuring a reusable, manufacturer-independent format\footnote{cf. Deliverable D2.16 of the EU H2020 project mySMARTLife.}.
This allows data consumers for an easy integration of the open data into their systems and applications.
The Urban Platform uses standard WebGIS Server Software to fulfil these requirements. 
This makes the SensorThings API a suitable choice to handle and provide real time (spatial) data.
These standard APIs are based on REST and SOAP with XML and JSON data formats using HTTP or MQTT.

\section{Field tests and experiments}
\label{sec:trials}

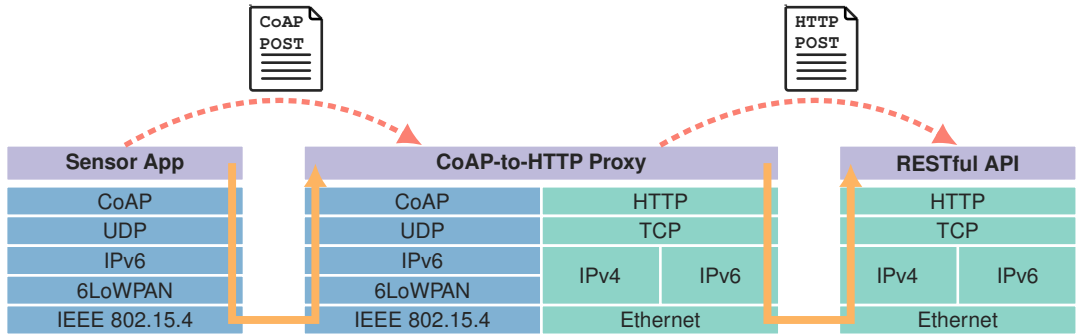
\begin{figure*}[t]
	\centering
\definecolor{DarkGrey}{RGB}{37,37,37}		
\definecolor{LiteGrey}{RGB}{217,217,217} 	
\definecolor{StackHTTP}{RGB}{141,211,199} 	
\definecolor{StackCoAP}{RGB}{128,177,211} 	
\definecolor{StackLink}{RGB}{253,180,98}		
\definecolor{App}{RGB}{190,186,218}			
\definecolor{AppLink}{RGB}{251,128,114}		

\tikzset{
	stack/.style = {
		draw=white, 
		rectangle, 
		line width=1pt, 
		font=\footnotesize\sffamily,
		text=DarkGrey,
		minimum width=20ex,
		minimum height=2.5ex,
		outer sep=0pt,
		inner sep=0pt,
		node distance=0pt,
	},
	http/.style 	 	= {fill=StackHTTP},
	coap/.style 		= {fill=StackCoAP},
	app/.style			= {fill=App, minimum height=3ex,font=\footnotesize\sffamily\bfseries},
}
\tikzset{
	coapstack/.pic = {
		\node [stack, coap] (coap) {CoAP};
		\node [stack, coap, below=of coap.south] (udp) {UDP};
		\node [stack, coap, below=of udp.south] (ipv6) {IPv6};
		\node [stack, coap, below=of ipv6.south] (6lo) {6LoWPAN};
		\node [stack, coap, below=of 6lo.south] (mac) {IEEE 802.15.4};
		\begin{scope}
			\node[fit=(coap.north) (mac.south) (udp.east) (udp.west), inner sep = 0pt] (-all) {};
		\end{scope}
	},
	httpstack/.pic = {
		\node [stack, http] (http) {HTTP};
		\node [stack, http, below=of http.south] (tcp) {TCP};
		\node [stack, http, below=of tcp.south, minimum height=5ex,
			   rectangle split, rectangle split horizontal, rectangle split parts=2] (ip) 
			   {\nodepart[text centered, text width=10ex-0.5pt]{one}IPv4 
			    \nodepart[text centered, text width=10ex-0.5pt]{two}IPv6};
		\node [stack, http, below=of ip.south] (mac) {Ethernet};
		\begin{scope}
			\node[fit=(http.north) (mac.south) (tcp.east) (tcp.west), inner sep = 0pt] (-all) {};
		\end{scope}
	},
	restmsg/.pic = {
		\def\w{6ex};
		\def\h{7ex};
		\def\c{1ex};
		\def\r{1pt};
		\def\lw{1pt};
		\def\ln{4};
		\coordinate (nw) at (-1ex,-3ex);
		\coordinate (ne0) at ($(nw) + (\w, 0)$);
		\coordinate (ne1) at ($(ne0) - (\c, 0)$);
		\coordinate (ne2) at ($(ne0) - (0, \c)$);
		\coordinate (se) at ($(ne0) + (0, -\h)$);
		\filldraw [-, DarkGrey, line width = \lw, fill=white] (nw) -- (ne1) -- (ne2)
				  [rounded corners=\r] -- (se) -- (nw|-se) -- (nw) -- cycle;
     	\draw [-, line width = \lw] (ne1) [rounded corners=\r]-- (ne1|-ne2) -- (ne2);
     	\node [anchor=north west, text width=4ex, text=DarkGrey] at (nw) {\scriptsize\ttfamily\bfseries\tikzpictext\par};
     	\foreach \k in {1,...,\ln}
     	{
       		\draw [-, DarkGrey, line width = \lw, line cap=round] 
         	($(nw|-se) + (1ex,1ex) + (0,{(\k-1)/(\ln-1)*(\h - 5ex)})$)
           	-- ++ ($(\w,0) - (2ex,0)$);
     	}
	},
}
\tikzsetnextfilename{crosscoap}
\begin{tikzpicture}
	\node [stack, app] (s_app) at (0,0) {Sensor App};
	\pic[draw, below = 2pt of s_app.south] (stack1) {coapstack};
	\node [stack, minimum width = 40ex, app, right=5ex of s_app] (crosscoap) {CoAP-to-HTTP Proxy};
	\pic[draw, below left=2pt and 0pt of crosscoap.south] (stack2) {coapstack};
	\pic[draw, below right=2pt and 0pt of crosscoap.south] (stack3) {httpstack};
	\node [stack, app, right=5ex of crosscoap] (backend) {RESTful API};
	\pic[draw, below = 2pt of backend.south] (stack4) {httpstack};
	\draw[-{Latex[length=2ex,width=2ex]}, line width=3pt, color=StackLink] 
		([xshift=-1ex]s_app.east) -- 
		([xshift=-1ex]stack1mac.east) -- 
		([xshift= 1ex]stack2mac.west) --
		([xshift= 1ex]crosscoap.west) {};
	\draw[-{Latex[length=2ex,width=2ex]}, line width=3pt, color=StackLink] 
		([xshift=-1ex]crosscoap.east) -- 
		([xshift=-1ex]stack3mac.east) -- 
		([xshift= 1ex]stack4mac.west) --
		([xshift= 1ex]backend.west) {};
	\draw[-{Latex[length=2ex,width=2ex]}, line width=2pt, color=AppLink, bend left, densely dashed] (s_app.north) to node [auto] (coaplink) {} ($(crosscoap.north) - (10ex,0)$); 
	\draw[-{Latex[length=2ex,width=2ex]}, line width=2pt, color=AppLink, bend left, densely dashed] ($(crosscoap.north) + (10ex,0)$) to node [auto] (httplink) {} (backend.north);
	\pic[above=10ex of coaplink.west, pic text = {CoAP POST}] (coapmsg) {restmsg};
	\pic[above=10ex of httplink.west, pic text = {HTTP POST}] (httpmsg) {restmsg};
\end{tikzpicture}
	\caption{End-to-end data flow of a vertically integrated IoT sensor application transmitting measurement data to an OGC Sensorthings RESTful API backend.}
	\label{fig:crosscoap}
\end{figure*}

Parallel to the design process of the MONICA IoT platform architecture, we started at an early stage to implement and deploy experimental IoT setups in Hamburg.
The motivation behind this was to examine and verify IoT protocols and technologies for usage within the MONICA project in real-world scenarios, and how to realise integration of the Hamburg Urban Platform and the MONICA platform, later on.

\subsection{Overview on Setup and Deployment}
\label{ssec:deploy}

The deployment consists of 3 components: (a) an IoT sensor node based on RIOT-OS, (b) a simple gateway node running a CoAP-to-HTTP proxy software, and (c) a HTTP server providing a RESTful API backend that conforms to the OGC SensorThings specification.

Figure~\ref{fig:crosscoap} shows the end-to-end data flow as well as the specific network protocols involved.
On the left-hand side is the sensor application that is based on RIOT-OS and runs on an embedded, constrained IoT node with a low-power radio interface (namely IEEE 802.15.4).
For this reason a network protocol stack specifically tailored for constrained environments is used, the stack consists of CoAP, UDP, and IPv6 with the 6LoWPAN adaption layer.

The JSON encoded sensor data is send (pushed) using a CoAP POST request to the RESTful API backend server  via a gateway node, which runs a CoAP-to-HTTP proxy software.
From an end-to-end perspective on the application layer the gateway node is a (nearly) transparent proxy, but is needed for two reasons:
(a) forward data from constrained network to the Internet, and (b) provide proxy functionality to translate between CoAP and HTTP.

It should be noted that in this setup the IoT node has an IPv6 address assigned, hence it would also be possible to receive (pull) data directly from the IoT sensor node using a RESTful GET request.
Physical network communication between the sensor and gateway node was realised using a wireless, low-power IEEE 802.15.4 radio transceiver.
Both nodes were deployed within close proximity to establish a direct (one-hop) wireless communication link.
The gateway forwards sensor data over the Internet to the remote backend server (cloud) using a wired ethernet connection.

\subsection{OGC SensorThings Integration}
\label{ssec:ogc}

For the integration of sensor and real time data to the Hamburg Urban Platform an implementation of the OGC SensorThings API based on the \emph{Fraunhofer Open Source SensorThings API Server} (FROST Server)\footnote{\url{https://github.com/FraunhoferIOSB/SensorThingsServer}} is used.
Before sending any data from the IoT sensor node to the SensorThingsServer, we setup endpoints and handles that conform to the OGC specification as follows:
(a) a \emph{Thing}, representing the IoT node (Listing~\ref{lst:thing}), 
(b) a \emph{Sensor}, describing the sensor hardware used, 
(c) a \emph{Property}, specifying what is measured, and 
(d) a \emph{Datastream}, representing the data endpoint.

\begin{lstlisting}[caption={Definition of an IoT node as an OGC Thing in JSON.},label=lst:thing]
{
	"name": "RIOT Alpha",
	"description": "IoT sensor node",
	"properties": {
		"owner": "iNET RG, HAW Hamburg",
		"device": "Phytec phyNODE",
		"operating system": "RIOT-OS"
	},
	"Locations": [{
		"name": "BT7-R580A",
		"description": "Office",
		"encodingType": "application/vnd.geo+json",
		"location": {
			"type": "Point",
			"coordinates": [10.022993, 53.557189]
		}
	}]
}
\end{lstlisting}
%
%
%

The sensor application periodically (every 10s) sends temperature values (observations) to the OGC SensorThingsServer using the URL of the OGC datastream endpoint.
As the latter is linked to all required metadata, i.e., Thing, ObservedProperty, and Sensor, which completely describe a sensor reading (observation), the JSON encoded payload of the CoAP/HTTP POST request is relatively short: \texttt{\{"result":2143\}}\footnote{The \texttt{phenomenon time} is omitted to further reduce the payload size.}.
This results in small message sizes per observation, which in turn reduces the required bandwidth and avoids retransmissions due to collisions when using wireless low-power radio.

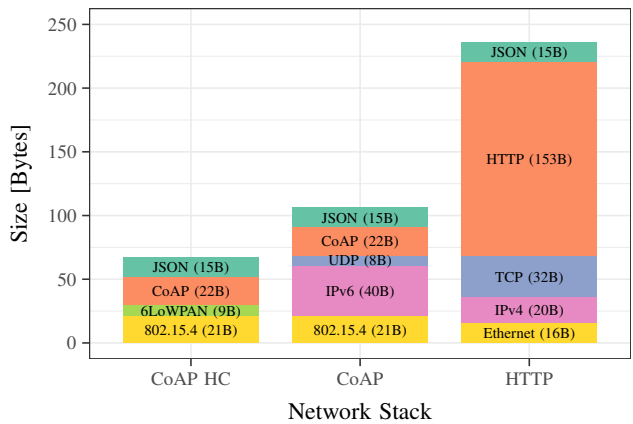
\begin{figure}[t]
	\centering
\begin{tikzpicture}[x=1pt,y=1pt]
\definecolor{fillColor}{RGB}{255,255,255}
\path[use as bounding box,fill=fillColor,fill opacity=0.00] (0,0) rectangle (245.72,166.22);
\begin{scope}
\path[clip] (  0.00,  0.00) rectangle (245.72,166.22);
\definecolor{drawColor}{RGB}{255,255,255}
\definecolor{fillColor}{RGB}{255,255,255}

\path[draw=drawColor,line width= 0.6pt,line join=round,line cap=round,fill=fillColor] (  0.00,  0.00) rectangle (245.72,166.22);
\end{scope}
\begin{scope}
\path[clip] ( 35.45, 28.08) rectangle (240.22,160.72);
\definecolor{fillColor}{RGB}{255,255,255}

\path[fill=fillColor] ( 35.45, 28.08) rectangle (240.22,160.72);
\definecolor{drawColor}{gray}{0.92}

\path[draw=drawColor,line width= 0.3pt,line join=round] ( 35.45, 46.17) --
	(240.22, 46.17);

\path[draw=drawColor,line width= 0.3pt,line join=round] ( 35.45, 70.28) --
	(240.22, 70.28);

\path[draw=drawColor,line width= 0.3pt,line join=round] ( 35.45, 94.40) --
	(240.22, 94.40);

\path[draw=drawColor,line width= 0.3pt,line join=round] ( 35.45,118.52) --
	(240.22,118.52);

\path[draw=drawColor,line width= 0.3pt,line join=round] ( 35.45,142.63) --
	(240.22,142.63);

\path[draw=drawColor,line width= 0.6pt,line join=round] ( 35.45, 34.11) --
	(240.22, 34.11);

\path[draw=drawColor,line width= 0.6pt,line join=round] ( 35.45, 58.23) --
	(240.22, 58.23);

\path[draw=drawColor,line width= 0.6pt,line join=round] ( 35.45, 82.34) --
	(240.22, 82.34);

\path[draw=drawColor,line width= 0.6pt,line join=round] ( 35.45,106.46) --
	(240.22,106.46);

\path[draw=drawColor,line width= 0.6pt,line join=round] ( 35.45,130.58) --
	(240.22,130.58);

\path[draw=drawColor,line width= 0.6pt,line join=round] ( 35.45,154.69) --
	(240.22,154.69);

\path[draw=drawColor,line width= 0.6pt,line join=round] ( 73.84, 28.08) --
	( 73.84,160.72);

\path[draw=drawColor,line width= 0.6pt,line join=round] (137.83, 28.08) --
	(137.83,160.72);

\path[draw=drawColor,line width= 0.6pt,line join=round] (201.82, 28.08) --
	(201.82,160.72);
\definecolor{fillColor}{RGB}{255,217,47}

\path[fill=fillColor] ( 48.24, 34.11) rectangle ( 99.44, 44.24);
\definecolor{fillColor}{RGB}{166,216,84}

\path[fill=fillColor] ( 48.24, 44.24) rectangle ( 99.44, 48.58);
\definecolor{fillColor}{RGB}{252,141,98}

\path[fill=fillColor] ( 48.24, 48.58) rectangle ( 99.44, 59.19);
\definecolor{fillColor}{RGB}{102,194,165}

\path[fill=fillColor] ( 48.24, 59.19) rectangle ( 99.44, 66.42);
\definecolor{fillColor}{RGB}{255,217,47}

\path[fill=fillColor] (112.24, 34.11) rectangle (163.43, 44.24);
\definecolor{fillColor}{RGB}{231,138,195}

\path[fill=fillColor] (112.24, 44.24) rectangle (163.43, 63.53);
\definecolor{fillColor}{RGB}{141,160,203}

\path[fill=fillColor] (112.24, 63.53) rectangle (163.43, 67.39);
\definecolor{fillColor}{RGB}{252,141,98}

\path[fill=fillColor] (112.24, 67.39) rectangle (163.43, 78.00);
\definecolor{fillColor}{RGB}{102,194,165}

\path[fill=fillColor] (112.24, 78.00) rectangle (163.43, 85.24);
\definecolor{fillColor}{RGB}{255,217,47}

\path[fill=fillColor] (176.23, 34.11) rectangle (227.42, 41.83);
\definecolor{fillColor}{RGB}{231,138,195}

\path[fill=fillColor] (176.23, 41.83) rectangle (227.42, 51.47);
\definecolor{fillColor}{RGB}{141,160,203}

\path[fill=fillColor] (176.23, 51.47) rectangle (227.42, 66.91);
\definecolor{fillColor}{RGB}{252,141,98}

\path[fill=fillColor] (176.23, 66.91) rectangle (227.42,140.70);
\definecolor{fillColor}{RGB}{102,194,165}

\path[fill=fillColor] (176.23,140.70) rectangle (227.42,147.94);
\definecolor{drawColor}{RGB}{0,0,0}

\node[text=drawColor,anchor=base,inner sep=0pt, outer sep=0pt, scale=  0.57] at ( 73.84, 37.21) {802.15.4 (21B)};

\node[text=drawColor,anchor=base,inner sep=0pt, outer sep=0pt, scale=  0.57] at ( 73.84, 44.45) {6LoWPAN (9B)};

\node[text=drawColor,anchor=base,inner sep=0pt, outer sep=0pt, scale=  0.57] at ( 73.84, 51.92) {CoAP (22B)};

\node[text=drawColor,anchor=base,inner sep=0pt, outer sep=0pt, scale=  0.57] at ( 73.84, 60.85) {JSON (15B)};

\node[text=drawColor,anchor=base,inner sep=0pt, outer sep=0pt, scale=  0.57] at (137.83, 37.21) {802.15.4 (21B)};

\node[text=drawColor,anchor=base,inner sep=0pt, outer sep=0pt, scale=  0.57] at (137.83, 51.92) {IPv6 (40B)};

\node[text=drawColor,anchor=base,inner sep=0pt, outer sep=0pt, scale=  0.57] at (137.83, 63.50) {UDP (8B)};

\node[text=drawColor,anchor=base,inner sep=0pt, outer sep=0pt, scale=  0.57] at (137.83, 70.74) {CoAP (22B)};

\node[text=drawColor,anchor=base,inner sep=0pt, outer sep=0pt, scale=  0.57] at (137.83, 79.66) {JSON (15B)};

\node[text=drawColor,anchor=base,inner sep=0pt, outer sep=0pt, scale=  0.57] at (201.82, 36.01) {Ethernet (16B)};

\node[text=drawColor,anchor=base,inner sep=0pt, outer sep=0pt, scale=  0.57] at (201.82, 44.69) {IPv4 (20B)};

\node[text=drawColor,anchor=base,inner sep=0pt, outer sep=0pt, scale=  0.57] at (201.82, 57.23) {TCP (32B)};

\node[text=drawColor,anchor=base,inner sep=0pt, outer sep=0pt, scale=  0.57] at (201.82,101.85) {HTTP (153B)};

\node[text=drawColor,anchor=base,inner sep=0pt, outer sep=0pt, scale=  0.57] at (201.82,142.36) {JSON (15B)};
\definecolor{drawColor}{gray}{0.20}

\path[draw=drawColor,line width= 0.6pt,line join=round,line cap=round] ( 35.45, 28.08) rectangle (240.22,160.72);
\end{scope}
\begin{scope}
\path[clip] (  0.00,  0.00) rectangle (245.72,166.22);
\definecolor{drawColor}{gray}{0.30}

\node[text=drawColor,anchor=base east,inner sep=0pt, outer sep=0pt, scale=  0.72] at ( 30.50, 31.63) {0};

\node[text=drawColor,anchor=base east,inner sep=0pt, outer sep=0pt, scale=  0.72] at ( 30.50, 55.75) {50};

\node[text=drawColor,anchor=base east,inner sep=0pt, outer sep=0pt, scale=  0.72] at ( 30.50, 79.86) {100};

\node[text=drawColor,anchor=base east,inner sep=0pt, outer sep=0pt, scale=  0.72] at ( 30.50,103.98) {150};

\node[text=drawColor,anchor=base east,inner sep=0pt, outer sep=0pt, scale=  0.72] at ( 30.50,128.10) {200};

\node[text=drawColor,anchor=base east,inner sep=0pt, outer sep=0pt, scale=  0.72] at ( 30.50,152.21) {250};
\end{scope}
\begin{scope}
\path[clip] (  0.00,  0.00) rectangle (245.72,166.22);
\definecolor{drawColor}{gray}{0.20}

\path[draw=drawColor,line width= 0.6pt,line join=round] ( 32.70, 34.11) --
	( 35.45, 34.11);

\path[draw=drawColor,line width= 0.6pt,line join=round] ( 32.70, 58.23) --
	( 35.45, 58.23);

\path[draw=drawColor,line width= 0.6pt,line join=round] ( 32.70, 82.34) --
	( 35.45, 82.34);

\path[draw=drawColor,line width= 0.6pt,line join=round] ( 32.70,106.46) --
	( 35.45,106.46);

\path[draw=drawColor,line width= 0.6pt,line join=round] ( 32.70,130.58) --
	( 35.45,130.58);

\path[draw=drawColor,line width= 0.6pt,line join=round] ( 32.70,154.69) --
	( 35.45,154.69);
\end{scope}
\begin{scope}
\path[clip] (  0.00,  0.00) rectangle (245.72,166.22);
\definecolor{drawColor}{gray}{0.20}

\path[draw=drawColor,line width= 0.6pt,line join=round] ( 73.84, 25.33) --
	( 73.84, 28.08);

\path[draw=drawColor,line width= 0.6pt,line join=round] (137.83, 25.33) --
	(137.83, 28.08);

\path[draw=drawColor,line width= 0.6pt,line join=round] (201.82, 25.33) --
	(201.82, 28.08);
\end{scope}
\begin{scope}
\path[clip] (  0.00,  0.00) rectangle (245.72,166.22);
\definecolor{drawColor}{gray}{0.30}

\node[text=drawColor,anchor=base,inner sep=0pt, outer sep=0pt, scale=  0.72] at ( 73.84, 18.17) {CoAP HC};

\node[text=drawColor,anchor=base,inner sep=0pt, outer sep=0pt, scale=  0.72] at (137.83, 18.17) {CoAP};

\node[text=drawColor,anchor=base,inner sep=0pt, outer sep=0pt, scale=  0.72] at (201.82, 18.17) {HTTP};
\end{scope}
\begin{scope}
\path[clip] (  0.00,  0.00) rectangle (245.72,166.22);
\definecolor{drawColor}{RGB}{0,0,0}

\node[text=drawColor,anchor=base,inner sep=0pt, outer sep=0pt, scale=  0.90] at (137.83,  5.50) {Network Stack};
\end{scope}
\begin{scope}
\path[clip] (  0.00,  0.00) rectangle (245.72,166.22);
\definecolor{drawColor}{RGB}{0,0,0}

\node[text=drawColor,rotate= 90.00,anchor=base,inner sep=0pt, outer sep=0pt, scale=  0.90] at ( 11.70, 94.40) {Size [Bytes]};
\end{scope}
\end{tikzpicture}
	\caption{Size comparison of a RESTful POST message using CoAP and HTTP. As CoAP utilises IEEE 802.15.4 and 6LoWPAN with header compression (HC), the actual message size transmitted is smaller than the (uncompressed) message send by the sensor application, compare CoAP~HC and CoAP.}
	\label{fig:coap_http}
\end{figure}

Fig.~\ref{fig:coap_http} shows a comparison of the message size for a RESTful POST request using an IoT network stack with CoAP and a standard Internet stack with HTTP.
The total message sizes to transmit a JSON encoded sensor value as described above is 236 Bytes for HTTP compared to 67 Bytes for CoAP with 6LoWPAN and UDP header compression (HC), that means the HTTP POST is more than 3.5 times larger.
The uncompressed CoAP message ---as send by the sensor application--- has a size of 106 Bytes.
Header compression is applied within the network stack, this process is completely transparent to any application.
While the size of the JSON encoded sensor data (message payload) is 15 Bytes in all cases, the major difference in size is caused by HTTP which is 153\,B compared to 22\,B for CoAP.
It is noteworthy, that in our test deployment the CoAP-to-HTTP proxy also translates from IPv6 to IPv4 when passing data from the IoT network to the Internet, saving 20\,B for HTTP.
As of April 2018 we collected more than 2.8 million observations over a period of 11 months, while the experiment is still running.

\section{Conclusion and Outlook}
\label{sec:outlook}

In this paper, we introduced the EU project MONICA and its platform architecture for large-scale IoT deployments.
Further, we gave an overview on Smart City activities in Hamburg, which is a MONICA pilot site.
We reported on first field tests with an end-to-end sensor application that integrates dedicated IoT technologies and utilises a wide range of open standards and protocols.
Our work is based on the IoT operating system RIOT which has been already established in industrial IoT settings \cite{gkslp-inii-17}. 
It deploys a tailored network stack for constraint environments based on CoAP, UDP, IPV6 + 6LoWPAN, and IEEE 802.15.4, and a RESTful backend service based on the OGC SensorThings API.
In this experimental setup, we showed how to connect a constraint IoT network with the common Internet by using a CoAP-to-HTTP proxy to enable transparent, end-to-end data flow from an embedded IoT sensor to the Hamburg Urban Platform.
 
These field tests and experiments are ongoing work, and will be continued and updated iteratively in accordance to the progress within the MONICA project.
For 2018, we also plan to deploy other IoT application together with MONICA partners.
We also work on extending the current test setup by adding a variety of different IoT nodes and collect additional sensor readings.
With that the IoT network will change from a homogeneous, single-hop to a heterogeneous, multi-hop topology, hence requiring a routing protocol such as RPL.
This setup will also allow for experiments to study, measure, and compare the performance of IoT network protocols such as CoAP and MQTT in a large real-world deployment.
We aim to establish this extended test setup on an upcoming Hamburger DOM funfair in 2018.

\section*{Acknowledgment}

This work is co-funded by the European Union (EU) within the MONICA project under grant agreement number 732350. 
The MONICA project is part of the EU Framework Programme for Research and Innovation \emph{Horizon 2020} and the IoT European Large-Scale Pilots Programme.

\bibliographystyle{IEEEtran}
\bibliography{own,internet,local,rfcs,ids}

\end{document}